\documentclass[twocolumn]{aastex61}

\usepackage{amsmath}

\usepackage{xcolor}
\definecolor{darkgreen}{rgb}{0., 0.5, 0.}

\definecolor{darkred}{rgb}{0.6, 0., 0.}

\definecolor{blue}{rgb}{0.0, 0., 0.6}

\shorttitle{Collisions of NS with PBH as FRB engines}
\shortauthors{Abramowicz, Bejger and Wielgus}

\begin{document}

\title{Collisions of neutron stars with primordial black holes\\
as fast radio bursts engines}

\correspondingauthor{Micha{\l} Bejger}
\email{marek.abramowicz@physics.gu.se}
\author{Marek A. Abramowicz}

\affiliation{Physics Department, University of Gothenburg, 412-96 G{\"o}teborg, Sweden}
\affiliation{Nicolaus Copernicus Astronomical Center, Polish Academy of Sciences, 00-716, Warsaw, Poland}
\affiliation{Physics Department, Silesian University of Opava, Czech Republic} 

\author{Micha{\l} Bejger}
\email{bejger@camk.edu.pl}
\affiliation{APC, AstroParticule et Cosmologie, Universit{\'e} Paris Diderot, CNRS/IN2P3, CEA/Irfu, Observatoire de Paris, Sorbonne Paris Cit{\'e}, F-75205 Paris Cedex 13, France}
\affiliation{Nicolaus Copernicus Astronomical Center, Polish Academy of Sciences, 00-716, Warsaw, Poland}

\author{Maciek Wielgus}
\email{mwielgus@cfa.harvard.edu}
\affiliation{Black Hole Initiative, Harvard University, 20 Garden Street, Cambridge, MA 02138, USA}

\begin{abstract}
If primordial black holes with masses of $10^{25}\,\mbox{g}\gtrsim m \gtrsim
10^{17}\,\mbox{g}$ constitute a non-negligible fraction of the galactic dark-matter haloes, 
their existence should have observable consequences: 
they necessarily collide with galactic neutron stars, nest in
their centers and accrete the dense matter, eventually converting them 
to neutron-star mass black holes while releasing the neutron-star magnetic field 
energy. Such processes may explain the fast radio
bursts phenomenology, in particular their millisecond durations, large
luminosities ${\sim}10^{43}$ erg/s, high rate of occurrence $\gtrsim
1000/\mbox{day}$, as well as high brightness temperatures, 
polarized emission and Faraday rotation. Longer than the dynamical 
timescale of the Bondi-like accretion for light primordial black holes 
allows for the repeating fast radio bursts. 
This explanation follows naturally from (assumed) existence of the dark matter
primordial black holes and requires no additional unusual phenomena, in
particular no unacceptably large magnetic fields of neutron stars. In our
model, the observed rate of fast radio bursts throughout the Universe follows
from the presently known number of neutron stars in the Galaxy. 
\end{abstract}


%%%%%%%%%%%%%%%%%%%%%%%%%%%%%%%%%%%%%%%%
\section{Introduction} \label{sec:intro}
%%%%%%%%%%%%%%%%%%%%%%%%%%%%%%%%%%%%%%%%
Fast Radio Bursts (FRB) are one of the most ambiguous astronomical phenomena. 
The first FRB was discovered in archived data taken in 2001
by the Parkes radio telescope as a 30 Jy, heavily dispersed burst of 5 ms
duration with no clear connection to a host galaxy \citep{Lorimer2007}. To
date, a few tens of FRB were detected (for the up-to-date reference, see the
Swinburne Pulsar Group FRB catalogue, \citealt{Petroff2016}). 
So far a multitude of models was proposed to explain the FRB, ranging from the cataclysmic 
events involving mergers of neutron stars (NS) with other NS or black holes (BH), 
magnetized NS collapses, NS-quakes, active galactic nuclei, accretion processes 
in the vicinity of BH, to soft gamma repeaters and processes occurring in pulsars' magnetospheres 
(see \citealt{Katz2018} for a recent review regarding the proposed FRB models and caveats). 

From an observational point of view,  
FRB are characterized by millisecond durations, large radio luminosities, high
occurrence rates, and large dispersion measures suggesting their extragalactic
origin, with distances of the order of a Gpc. Their extremely high brightness
temperatures suggest the presence of plasma and/or magnetic fields \citep{Katz2014}. 
In at least one FRB source the event is (non-periodically)
repeated (FRB 121102, \citealt{Spilter2016}), and another one (FRB 121002,
\citealt{Champion2016}) shows a clear two-component peak profile. 
For FRB121102, recent observations show that pulse of radio waves 
passes through a veil of magnetized plasma \citep{Michilli2018}, resulting in 
detectable Faraday rotation. Currently, the main difficulty in selecting the `true' FRB model 
lies in the ambiguity of the available scant data, and in the problem of reconciliation 
the singular nature of the vast majority of the FRB with the repeating FRB121102.  

To us, the above feature list suggest a NS origin of the FRB engine. 
In the three following Sections, we show that the most puzzling FRB features, namely  
%-----------------------------------------------------------------------
\begin{itemize} 
\itemsep -0pt
\item FRB occurrence rate, ${\dot n}_{\rm obs} {\sim}10^3\ \mbox{1/day}$ (Sect.~\ref{sect:rate}), 
\item FRB luminosities, $L_{\rm obs} {\sim}10^{43}\ \mbox{erg/s}$ (Sect.~\ref{sect:energetics}) 
\item FRB duration times, $\delta t_{\rm obs} {\sim}10^{-3}\ \mbox{s}$ (Sect.~\ref{sect:duration}),
\item FRB repeaters (Sect.~\ref{sect:energetics} and \ref{sect:discussion}),
\end{itemize}  
%-----------------------------------------------------------------------
may be naturally explained by adopting one single assumption: that the FRB phenomenon is due to collisions of NS with primordial black holes (PBH) that are one of the constituents of the galactic dark matter haloes.  

Specifically, we show that during a PBH-NS collision, the PBH kinetic energy is
dissipated via the gravitational drag. This energy loss bounds the PBH to NS.
Eventually, the PBH settles at the NS center and accretes its material. 
This causes a NS to turn into a light, ${\sim}1.5 M_{\odot}$ BH,
in accretion timescales depending on the initial PBH mass. Magnetic energy
released during this process powers the FBR event. 
 
%%%%%%%%%%%%%%%%%%%%%%%%%%%%%%%%%%%%%%%%%%%
\section{FRB occurrence rate} \label{sect:rate} 
%%%%%%%%%%%%%%%%%%%%%%%%%%%%%%%%%%%%%%%%%%%

In the following we will discuss the rate of the FRB events, using simple order-of-magnitude
estimations, and the numerical simulations of the PBH halo interacting with a typical spiral galaxy.

%%%%%%%%%%%%%%%%%%%%%%%%%%%%%%%%%%%%%%%%%%%
\subsection{Analytic estimate} \label{subsect:rate-analytic} 
%%%%%%%%%%%%%%%%%%%%%%%%%%%%%%%%%%%%%%%%%%%
We argue that collisions of hypothetical PBH in the mass range
%-------------------------------------------------------------------------------------------------------
\begin{equation}
10^{17}\,\mbox{g} < m < 10^{25}\,\mbox{g}
\label{mass-range}
\end{equation}
with NS may explain the FRB phenomenology. This mass range is currently weakly explored by the astronomical observations \citep{Carr2016}, in principle allowing for the PBH to constitute a fraction of the galactic dark matter. We will adopt the values of $m$ from this mass range and to show that they are consistent with the results obtained. In the description of our model below, we will adopt a ``reference'' PBH mass from the unconstrained range of masses, 
%-------------------------------------------------------------------------------------------------------
\begin{equation}
m_0 = 10^{23}\ \mbox{g},
\label{reference-mass}
\end{equation}
%-------------------------------------------------------------------------------------------------------  
in order to be numerically specific. 

\citet{Abramowicz2009} calculated the number of the PBH-NS collisions that
occur in one day in a single galaxy (see their Table 3) and found that the
number scales inversely with the PBH mass:   
%-------------------------------------------------------------------------------------------------------
\begin{equation}
{\dot n} \sim 10^{-8} \left( \frac{m}{m_0} \right)^{-1}\, \mbox{1/day}.
\label{frequency-single}
\end{equation}
%-------------------------------------------------------------------------------------------------------
The repeating FRB 121102 \citep{Spilter2016} observed by the Arecibo telescope has been
identified with the host galaxy at $z=0.193$ ($\simeq 1$\ Gpc distance).
According to the FRB catalogue \citep{Petroff2016}, a number of FRB detected by
the Parkes telescope have the dispersion measure several times higher than the
FRB 121102, which suggests they are located at at least a few Gpc distances. 
Here we assume the FRB are detected from a large fraction of all the galaxies 
in the observable Universe $N\simeq 10^{11}$. One should multiply the Eq.~\ref{frequency-single} 
by $N$ to get an estimate of the rate of the FRB detections,
%-------------------------------------------------------------------------------------------------------
\begin{equation}
{\dot n}_{\rm model} \sim 10^{3} \left( \frac{m}{m_0} \right)^{-1}\, \mbox{1/day}.
\label{frequency-all-universe}
\end{equation}
%-------------------------------------------------------------------------------------------------------
This value agrees with that estimated from observations
\citep{Champion2016,Caleb2017}. In the next section we show that the occurrence
rate ${\sim}10^3\,\mbox{1/day}$ is consistent with the observed number of NS 
in the Galaxy. 

%%%%%%%%%%%%%%%%%%%%%%%%%%%%%%%%%%%%%%%%%%%
\subsection{Numerical simulations} \label{subsect:rate-numeric} 
%%%%%%%%%%%%%%%%%%%%%%%%%%%%%%%%%%%%%%%%%%%

The approximated shape of a galaxy is parametrized with 3 numbers, for which we assume following values:
%-------------------------------------------------------------------------------------------------------
\begin{figure}
	\includegraphics[width=\columnwidth]{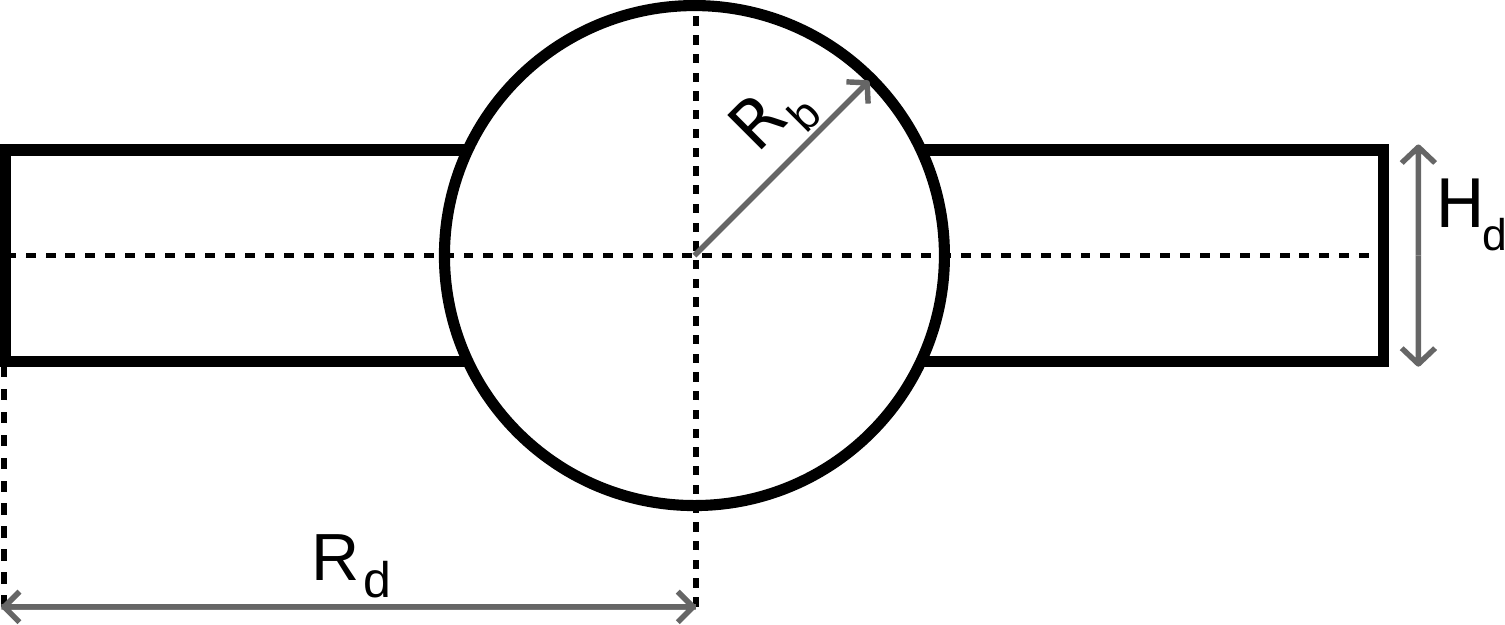} 
	\caption{The galaxy model: galactic disk thickness $H_{d} = 1.0 $ kpc, galactic disk radius $R_{d} = 30$ kpc, radius of the spherical galactic bulge $R_{b} = 5.0$ kpc.}
	\label{fig-galaxy}
\end{figure}
%-------------------------------------------------------------------------------------------------------
For the rotation of the galaxy we adopt a simplified rotational curve:  
%------------------------------------------------------------------------------------------------------- 
\begin{align}
  V(r) =
  \begin{cases}
    v_d = 220 \ \text{km/s,}\quad \text{for}\ r > R_b,\\    
    \frac{v_d}{R_b} r\ \text{km/s,}\quad \text{for}\ r \le R_b,        
  \end{cases}
\label{eq_velocity}
\end{align}
%-------------------------------------------------------------------------------------------------------
which allows to calculate the total mass density from $G{\cal M}(r)/r^2 = V^2(r)/r$, where ${\cal M}(r)$ is the mass within the radius $r$. The density $\rho(r)$ within the radius $r$ is 
%-------------------------------------------------------------------------------------------------------
\begin{align}
\rho(r) = 
  \begin{cases}
    \frac{v_d^2}{4 \pi G r^2},\quad \text{for} \ r > R_b,\\ 
    \frac{3 v_d^2}{4 \pi G R_b^2},\quad \text{for}\ r \le R_b.  
  \end{cases} 
\label{eq_density} 
\end{align}
%-------------------------------------------------------------------------------------------------------
Total density remains unchanged during the simulation and mass of the galaxy is found by integration,
%-------------------------------------------------------------------------------------------------------
\begin{equation}
{\cal M}_{tot} = \frac{v_d^2}{G} \left[R_b + \frac{H_d}{2} \ln\left(\frac{R_d}{R_b} \right)  \right] \sim 7\times 10^{10} M_{\odot}.
\label{galaxy_mass}
\end{equation} 
%-------------------------------------------------------------------------------------------------------
In the model we consider evolution of four classes of astrophysical species,
listed in the Table~\ref{tab:1}. 
 
%-----------------------------------------------------------------------
\begin{table}[h]
\begin{tabular}{|l| l l l |}
\hline
(Index) Name       & Mass                        & $R_i/R_{G,i}$              & Number \\[-5pt]
                   & $M_i$                       & ratio                  & density $n_i$ \\  
\hline 
  (N) NS           & $M_N=1.5 M_{\odot}$         & 2.5                     & $n_N$      \\
  (P) PBH          & $ \langle M_P \rangle =\overline{m}$                  & 1                       & $n_P$      \\
  (0) Light BH     & $M_0=1.5 M_{\odot}$         & 1                       & $n_0$      \\
  (1) Stellar BH   & $M_1=10 M_{\odot}$          & 1                       & $n_1$      \\
\hline    
\end{tabular}
\label{tab:1} 
\caption{Main parameters of the astrophysical species adopted in the analysis: their masses $M_i$, radii $R_i$ and 
number densities $n_i$. The gravitational radius for each species is denoted by $R_{G,i}$.}
\end{table}
%-----------------------------------------------------------------------

Here $R_{G,i} = 2GM_i/c^2$ is the gravitational radius of a species with the
index $i$ and the mass $M_i$. The parameter $\overline{m}$ represents the mean value of the 
PBH mass distribution. Light BH result from the NS-PBH collisions and
stellar BH are the outcome of the standard stellar evolution. 

Evolution of the number densities $n_i$ of the four species result from interactions between them which is described by collision and creation coefficients, $C^{ij}$ [pc$^3$/y] and $K^{i}$  [pc$^{-3}$ y$^{-1}$], respectively. If mass of $i$-th element dominates over mass of $j$-th element, we assume a rate of collisions between species $i$ and $j$ to be:
%-------------------------------------------------------------------------------------------------------
\begin{equation}
C^{ij}n_i n_j = n_i n_j \frac{V_i^2}{V(r)}R_i^2 ,
\end{equation}
%-------------------------------------------------------------------------------------------------------
$V(r)$ is the rotational velocity given by Eq.~\ref{eq_velocity}, $V_i$ is the escape velocity characteristic to the i-th species, 
%-----------------------------------------------------------------
\begin{equation}
V_{i} \equiv v_* = \sqrt{\frac{2GM_i}{R_i }},    
\label{eq-escape-velocity} 
\end{equation}
%----------------------------------------------------------------- 
which in case of BH species leads to $V_P = V_0 = V_1 = c$.   
Creation operators aim at capturing the rate of NS and BH creation in standard stellar evolution processes. Therefore, we assume a total creation rate for the whole galaxy, $K^{i}_{tot}$, and weight it with a total local density $\rho(r)$,
%-------------------------------------------------------------------------------------------------------
\begin{equation}
K^i(r) = \frac{K^{i}_{tot}}{{\cal M}_{tot}}\rho(r) ,
\label{Kcreation}
\end{equation}
%-------------------------------------------------------------------------------------------------------
where ${\cal M}_{tot}$ is a total mass of the galaxy, Eq.~\ref{galaxy_mass}. Assumed total creation rates are $K^{N}_{tot} =K^{1}_{tot}  = 0.01$\,y$^{-1}$, reflecting the average supernova rate in typical galaxy. 
%-------------------------------------------------------------------------------------------------------
From this considerations it follows that the evolution of the four species number densities are given by a following set of differential equations:
%-------------------------------------------------------------------------------------------------------
\begin{align}
 %\label{eq_sys_main_1} 
\frac{\partial n_N}{\partial t} = & -C^{NP}n_N n_P   - C^{NN}n_N n_N - C^{N0}n_N n_0  \\ \label{eq_sys_main_1}
&- C^{N1} n_N n_1 + K^N , \nonumber \\ 
\frac{\partial n_P}{\partial t} = & -C^{NP} n_N n_P - C^{P0} n_P n_0 - C^{P1} n_P n_0 \\
&- C^{PP} n_P n_P , \nonumber \\ 
\frac{\partial n_0}{\partial t} =\ & C^{NP} n_N n_P  - C^{00} n_0 n_0 -C^{01} n_0 n_1 ,\\ 
\frac{\partial n_1}{\partial t} =\ & C^{NN} n_N n_N + K^1 . \label{eq_sys_main_4} \\ 
\nonumber
\label{eq_model}
\end{align}
%-----------------------------------------------------------------------
Both collision and creation coefficients depend on the location within a
galaxy. Hence, we solve the Eqs.~\ref{eq_sys_main_1}-\ref{eq_sys_main_4} separately
in multiple zones, characterized by their specific rotational velocity $V(r)$
and density $\rho(r)$. For the computational reasons we divide galactic disk
into 50 cylindrical zones, in which $\rho$ and $V$ were assumed to be constant.
Similarly, bulge region was divided into 50 spherical zones.

\begin{figure}
	\includegraphics[width=\columnwidth]{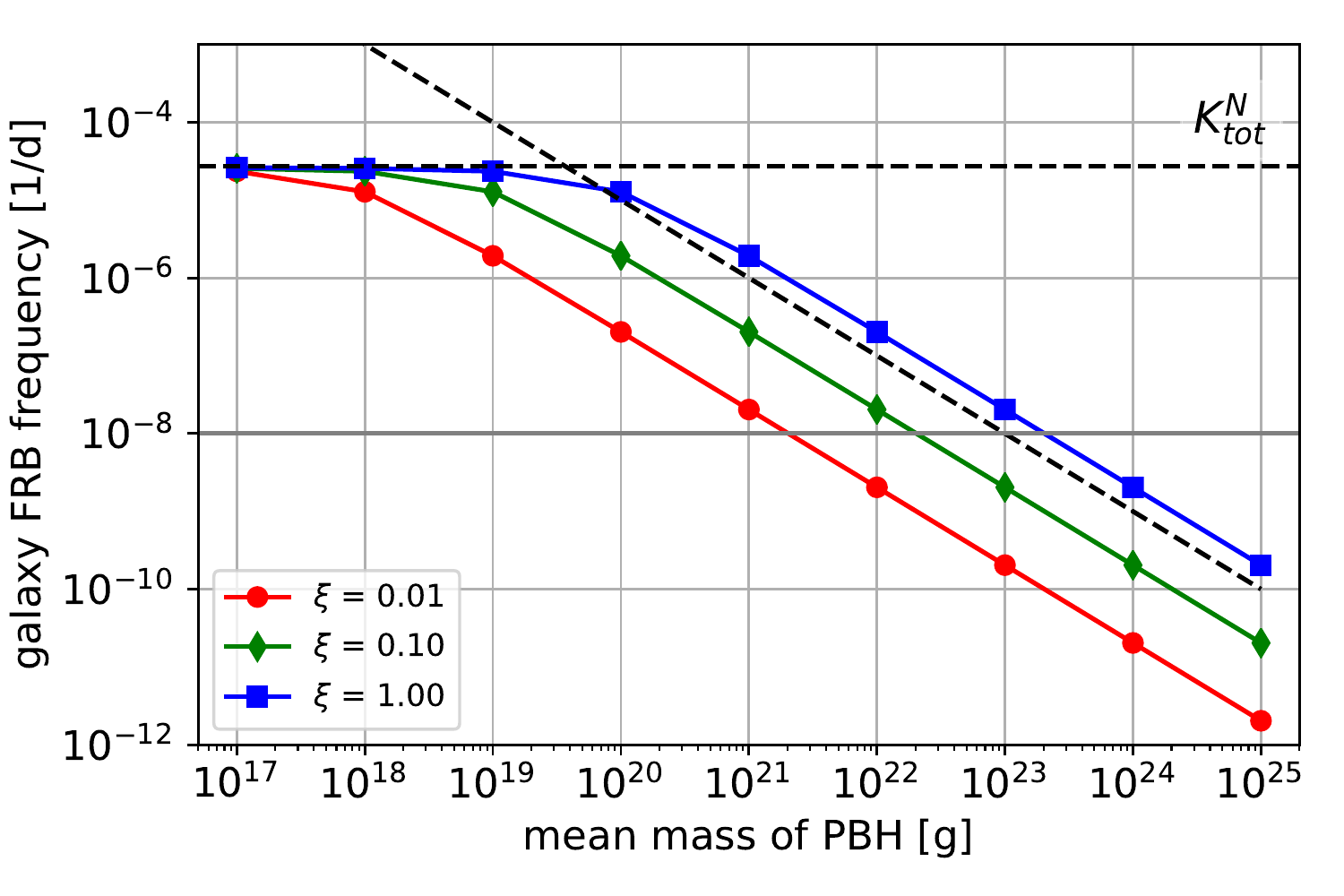} 
	\caption{Galactic rates of FRB (NS-PBH interactions) as function of mean PBH mass for three values of the ratio between PBH mass density and total gravitating density in the galaxy $\xi=1.0, 0.1, 0.01$. Obtained with the numerical simulation (continuous lines) and with Eq. \ref{frequency-single} (dashed line). Gray horizontal line denotes the frequency of $10^{-8}$ 1/day, necessary to explain the FRB occurrence rate.}
	\label{frb_galaxy_rate}
\end{figure}

In the simulation we fix the PBH mass $M_P = \overline{m}$. Nevertheless, for the presented
system of equations, as long as $M_P \ll M_N$, this is tantamount to having a distribution of PBH masses, with expected value $\overline{m}$. The equations are independent of the higher moments of the PBH mass distribution. The simulations were initialized with $n_N(0) = n_0(0) = n_1(0) = 0$. Initial
number density of PBH is zone-dependent,
%-------------------------------------------------------------------------------------------------------
\begin{equation}
n_P(t = 0, r) = \xi \frac{\rho(r)}{\overline{m}},
\end{equation}
%------------------------------------------------------------------------------------------------------- 
where $\xi$ represents the ratio between PBH mass density and total gravitating
density in the galaxy. Equations are then integrated for $T_{tot} = 1.3 \times
10^{10}$ y in parallel for all zones. Finally, a current total rate of FRB
events in a galaxy can be calculated by multiplying expressions $C^{NP}
n_N(T_{tot})n_P(T_{tot})$ times the respective zone volume and summing it up
over all zones. Results are shown in Fig.~\ref{frb_galaxy_rate} as function of $\overline{m}$. 
We have evaluated models with three values of $\xi$: $1.0, 0.1, 0.01$. 
Note that the value of $\xi=1.0$ is used here for comparison. We do not a priori 
assume that PBHs are the dominant ingredient of DM, but that PBHs may constitute 
a small fraction of DM.   
For larger $\xi$ results are consistent with a simple estimation given by
Eq.~\ref{frequency-single} for $\overline{m} > 10^{20}$~g. Models with lower $\xi$ 
predicts lower FRB frequency for the same mean PBH mass. For $\overline{m} < 10^{20}$~g numerical models
qualitatively deviate from Eq.~\ref{frequency-single}. In this regime a large
number density of light PBH is limiting the total number of NS. In the low $\overline{m}$ limit rate of FRB asymptotically reaches the NS creation rate $K^{N}_{tot}$, indicating equilibrium between number of NS created and destroyed in a unit time.

Parameters such as geometry of a model galaxy were found to have a limited influence on the results and may modify the galactic FRB frequency only by a factor of order of unity.
%-------------------------------------------------------------------------------------------------------
%-------------------------------------------------------------------------------------------------------
\begin{figure}
	\includegraphics[width=\columnwidth]{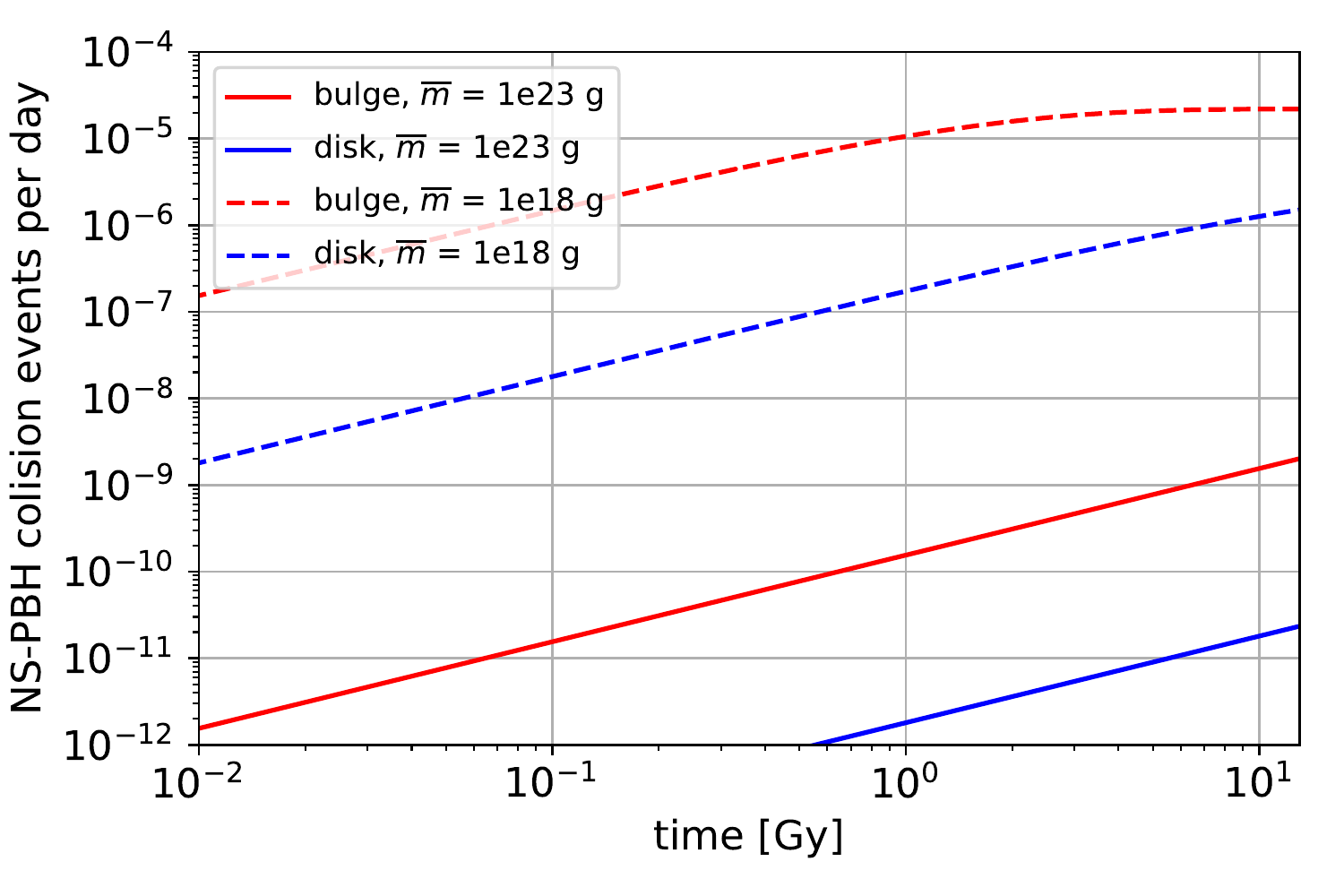} 
	\caption{Number of NS-PBH collisions in a galactic bulge and disk for different mean PBH masses and $\xi=0.1$, as function of the age of a galaxy.}
	\label{frb_galaxy_rate2}
\end{figure}
%-------------------------------------------------------------------------------------------------------
\begin{figure}
	\includegraphics[width=\columnwidth]{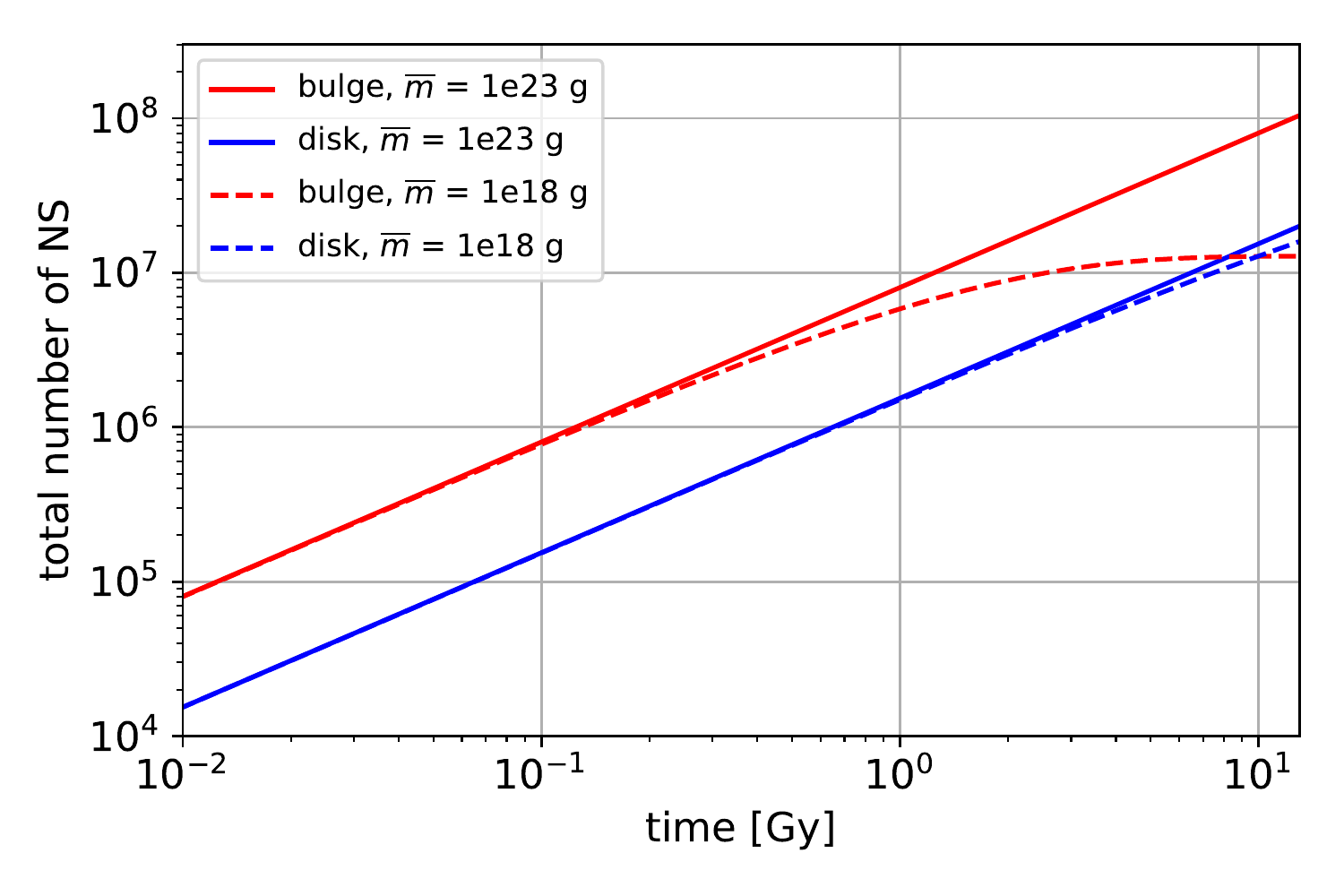} 
	\caption{Total number of NS in a galactic bulge and disk for different mean PBH masses and $\xi=0.1$, as function of the age of a galaxy.}
	\label{frb_galaxy_rate3}
\end{figure}
%-------------------------------------------------------------------------------------------------------

\citet{Pani2014} discusses constraints on the PBH fraction in dark matter in the relevant mass range, suggesting that $\xi$ of at most few percent should be adopted (see also \citealt{Carr2016} for similar conclusions). In our model this corresponds to a shift in mean PBH mass, necessary to explain the FRB occurrence rate, from about $10^{23}$ g for $\xi=1.00$ to about $10^{21}$ g for $\xi=0.01$, still within the acceptable mass range.   
Figure~\ref{frb_galaxy_rate2} shows the number of NS-PBH collisions in a
galactic bulge and disk for different mean PBH masses and selected value of
$\xi=0.1$, as function of the age of a galaxy. In general, there is far more
PBH-NS collisions in the bulge than in the disk. However, for small values of
$\overline{m}$ the number of collisions in the bulge saturates after ${\sim}10$\,Gy.
Therefore, this general trend reverses for small $\overline{m}$.
Figure~\ref{frb_galaxy_rate3} shows the total number of NS in a galactic bulge
and disk for different mean PBH masses and $\xi=0.1$, as function of the age of
a galaxy. Our simulation shows that the number of NS in the bulge saturates
with time because NS are destroyed by numerous PBH. Indeed, in Eq.~\ref{eq_sys_main_4} the dominant terms are the first one (local rate of NS-PBH collisions) and the last one (local NS creation rate). As first one grows with NS local number density $n_N$, and the latter is modeled as time-independent, Eq. \ref{Kcreation}, the saturation of NS number may be reached on a relevant timescale with $n_N \propto K^{N}/n_p \propto K^{N} \overline{m} $. This also means that if
$\overline{m}$ is small, then with increasing cosmological time, the total number of NS
in the disk surpasses the NS number in the bulge.

%%%%%%%%%%%%%%%%%%%%%%%%%%%%%%%%%%%%%%%%%%%
\section{FRB luminosities} \label{sect:energetics} 
%%%%%%%%%%%%%%%%%%%%%%%%%%%%%%%%%%%%%%%%%%%

In our model, the energetics of an individual FRB is quite universal, in the
sense that the energy output and the duration of the event follow from the typical 
parameters of the NS. 

%-%-%-%-%-%-%-%-%-%-%-%-%-%-%-%-%-%-%-%-%-%
\subsection{Energetics of the PBH passage through a NS} 
%-%-%-%-%-%-%-%-%-%-%-%-%-%-%-%-%-%-%-%-%-%

Close gravitational encounters of PBH and NS, their capturing and orbital
energy loss via e.g., the tidal interaction were extensively studied by
\citet{Capela2013,Defillon2014,Kouvaris2014,Pani2014}. To make our case here,
we provide simpler arguments to show that once a PBH interacts sufficiently
closely with the NS, fate of the latter is determined.  Let us assume that a
PBH directly collides with the NS \citep{Abramowicz2009}. The PBH-NS collisions occur
with velocity which at the NS surface is equal, or (slightly) higher than the
NS escape velocity (Eq.~\ref{eq-escape-velocity}). 

\begin{figure}
	\includegraphics[width=\columnwidth]{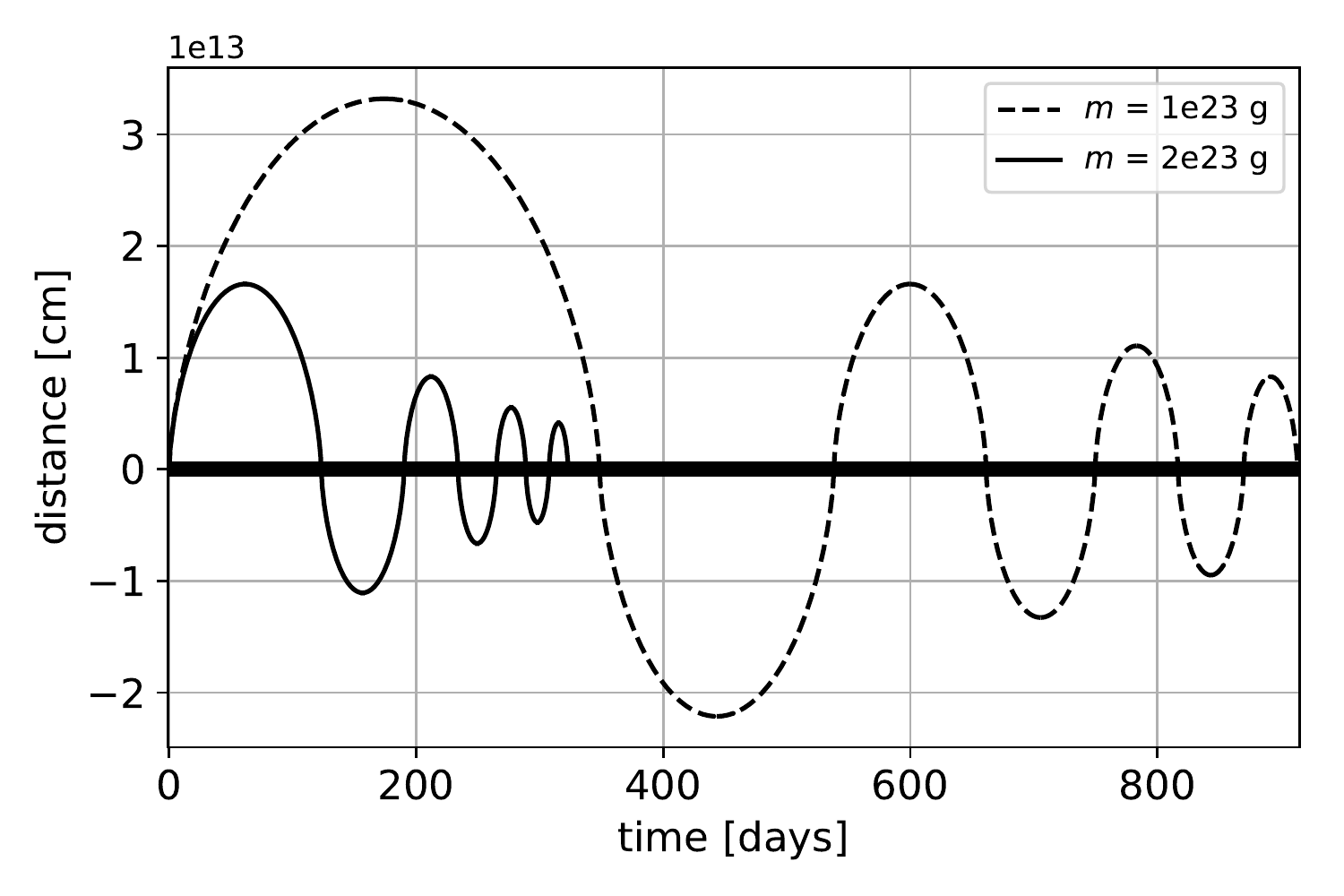} 
	\caption{Examples of PBH trajectories in NS gravitational field after the initial interaction. Each pass through NS reduces PBH's kinetic energy by $E_*$, Eq. \ref{eq-energy-losses}.} 
	\label{frb_kinetic_loss}
\end{figure}

During the passage through the NS, a PBH does not interact directly with
the NS matter, in particular it does not accrete. The strongest interaction is
only indirect -- via gravitational drag. When the PBH passes through the dense
matter gas, gravity transfers momentum to nuclei nearby. This focuses matter
behind the PBH and creates a wake that gives rise to a force acting on the PBH
and causes a decrease in its kinetic energy, $E_{\rm kin}$. A relevant case
(i.e. when the PBH velocity is highly supersonic) was studied by
\citet{Ruderman1971}, \citet{Ostriker1999}  and more recently by
\citet{Capela2013}. \citet{Abramowicz2009} employed these sort of studies to
calculate the kinetic energy losses when the PBH moves through an NS (see their
Table 3),
%-----------------------------------------------------------------
\begin{equation}
E_* = 3\times 10^{35} \left( \frac{m}{m_0} \right)^2 \mbox{erg}.
\label{eq-energy-losses} 
\end{equation}
%-----------------------------------------------------------------
After loosing energy $E_*$ a PBH will have, typically, insufficient kinetic
energy to escape to infinity after passing through the NS: it will turn at the
``escape radius'' $r_0$ and collide with the NS again. From the equation of
energy conservation, 
%-----------------------------------------------------------------
\begin{equation}
-\frac{GMm}{r_0} + {E_*} = 0, 
\label{eq-energy-conservario-radius-escape} 
\end{equation}
%-----------------------------------------------------------------
and, from Eqs.~\ref{eq-energy-losses}-\ref{eq-energy-conservario-radius-escape}, we derive for the escape radius,
%-----------------------------------------------------------------
\begin{equation}
r_0 \approx 10^{13} \left( \frac{m_0}{m} \right)\ \mbox{cm}.
\label{eq-radius-escape} 
\end{equation}
%-----------------------------------------------------------------
The gradual loss of PBH's kinetic energy as it nests inside a NS is illustrated in Fig. \ref{frb_kinetic_loss}.
%-------------------------------------------------------------------------------------------------------
%-------------------------------------------------------------------------------------------------------

``Magnetic drag'' extorted on a PBH during its passage through
NS is irrelevant even in the case of $B\,{\sim}10^{15}$\,G. Motion of a BH in
an external magnetic field was studied by many authors, including
\citet{Wald1974}, and more recently \citet{Morozova2014}. From this recent
study we adopt their Eqs.~(74--76) relevant to our model. It describes the PBH
kinetic energy losses due to its interaction with the NS magnetic field,
%-------------------------------------------------------------------------------------------------------
\begin{equation}
L \approx 10^{30} \left( \frac{m}{m_0}\right)^2 \left( \frac{B}{B_0} \right)^2\, \mbox{erg/s}, 
\label{kinetic-Wald}
\end{equation}
%-------------------------------------------------------------------------------------------------------
with $B_0 = 10^{15}$ G. This is far too small to be relevant for the FRB energetics for realistic internal magnetic fields of NS. 

If the PBH is spinning ($a=$ dimensionless spin) and the NS possesses a
substantial magnetic field in its interior, then one expects that the
Blandford-Znajek mechanism \citep{Blandford1977} will give the following 
power:  
 
%-------------------------------------------------------------------------------------------------------
\begin{equation}
L^{BZ} \sim 10^{29}\, a^2 \left( \frac{m}{m_0}\right)^2 \left( \frac{B}{B_0} \right)^2\, \mbox{erg/s},  
\label{Blandford-Znajek}
\end{equation}
%-------------------------------------------------------------------------------------------------------
see e.g., \citet{Lee2000} (their Eq.~D.29). This is again far too small 
to be relevant for the FRB energetics.

%-%-%-%-%-%-%-%-%-%-%-%-%-%-%-%-%-%-%-%-%-%
\subsection{Energy reservoir for the burst} 
%-%-%-%-%-%-%-%-%-%-%-%-%-%-%-%-%-%-%-%-%-%

From Eq.~\ref{eq-radius-escape} it follows that lighter PBH with masses $m <
10^{18}$ g have a chance to fly away for more than about 1 pc, i.e., a typical
distance between stars in a galaxy. All other PBH return, collide again,
subsequently settle at the NS center and start to accrete the dense NS matter.
The timescale of Bondi-like accretion onto a PBH and creation of a NS-mass BH
was estimated in several works, e.g., \citet{Giddings2008} and
\citet{Fuller2017}: 
%-----------------------------------------------------------------
\begin{equation}
\frac{dM_P(t)}{dt} \sim A G^2 M_P^2(t)\rho_{c}, 
\label{eq-bondi} 
\end{equation}
%-----------------------------------------------------------------
where $A$ is a constant of the order of 10, depending on the details of the
equation of state of dense matter. For a central density of the order of
$\rho_c=10^{15}$ g/cm$^{3}$, corresponding to NS of mass $M_N=1.5\ M_\odot$,
the NS$\to$BH conversion time is 
%-----------------------------------------------------------------                            
\begin{equation}       
T = 10^{-3}\left(\frac{m_0}{M_P}\right)\ \mbox{y},                                            
\label{eq-tconv}       
\end{equation}         
%----------------------------------------------------------------- 
which for $M_P=m_0$ is of the order of a few hours. In the works quoted
above it was convincingly argued that the accretion of the NS matter onto a PBH
is a Bondi-like one. For our model the Bondi-likeness would be a convenient
feature (as it allows a simple theoretical description), but it is by no means 
necessary. Instead, it is quite sufficient that the timescale 
in Eq.~\ref{eq-tconv} is, in a range of PBH masses, much longer than a
free-fall collapse timescale.  

The accretion ends with a newly-born, light BH is formed, while the FRB results 
from a reconnecting NS magnetosphere. From this point of view of the FRB, 
a scenario presented here is a version of the ``blitzar'' model
proposed by \citet{Falcke2014}, in which the magnetosphere of a collapsing NS
reconnects and provides a source of energy for the FRB (see also
\citealt{Fuller2015}, where the collapse is triggered by accumulation of the
dark-matter particles). 

%-%-%-%-%-%-%-%-%-%-%-%-%-%-%-%-%-%-%-%-%-%
\subsection{ The energetics of the final burst} 
%-%-%-%-%-%-%-%-%-%-%-%-%-%-%-%-%-%-%-%-%-%

According to our model, the FRBs do not originate during
the passage of a PBH through NS, but during the final
collapse of NS to a light BH. Due to the no hair conjecture 
\citep{Israel1968,Carter1971}, the final BH cannot have magnetic field. 
From this general principle, one gets an estimate of the energy
available for an FRB in the NS magnetic field.

A typical NS of radius $R$ and the surface dipole 
magnetic field of the order of $B=10^{12}$ G contains enough energy for multiple 
$10^{40}$ erg bursts ($10^{43}$ erg/s and a millisecond duration): 
%-----------------------------------------------------------------                            
\begin{equation}
  E_B = \frac{B^2}{8\pi}\left(\frac{4}{3}\pi R^3\right)\simeq 10^{41}\ \mathrm{erg}. 
\label{eq-eb}
\end{equation}
%-----------------------------------------------------------------  
Note again that our model does not necessarily require extremely large (i.e.
magnetar size, $B\,{\sim}10^{15}$\,G) magnetic fields. Magnetic fields play a
role only in the emission of radio waves.  

%-%-%-%-%-%-%-%-%-%-%-%-%-%-%-%-%-%-%-%-%-%
\subsection{Repeating nature of FRB121102} 
%-%-%-%-%-%-%-%-%-%-%-%-%-%-%-%-%-%-%-%-%-%

From the timescales estimated above, it is likely that for large enough $M_P$
most of the FRB are observed producing one, final burst. However, for a 
small-mass PBH, the accretion may take much longer time (of the order of 
years or more, for $M_P<10^{20}$ g). A long-lasting accretion scenario creates a
tantalizing opportunity to explain the repeating FRB 
- as the gradual accretion takes place in the superfluid, magnetized interior of 
the NS (see e.g., \citealt{Glampedakis2011}), occasional reconnection of the magnetic field 
line bundles confined in the superfluid may be responsible for irregular bursts of 
FRB121102 \citep{Oppermann2018}.

%-%-%-%-%-%-%-%-%-%-%-%-%-%-%-%-%-%-%-%-%-%
\subsection{Gravitational waves emission} 
%-%-%-%-%-%-%-%-%-%-%-%-%-%-%-%-%-%-%-%-%-%

The conversion of an NS into a light BH is also related to spinning up the object, 
and possibly also mass-shedding of some outermost NS layers (see \citealt{Fuller2017} 
in the context of the r-process elements production). 

\begin{figure}
	\includegraphics[width=\columnwidth]{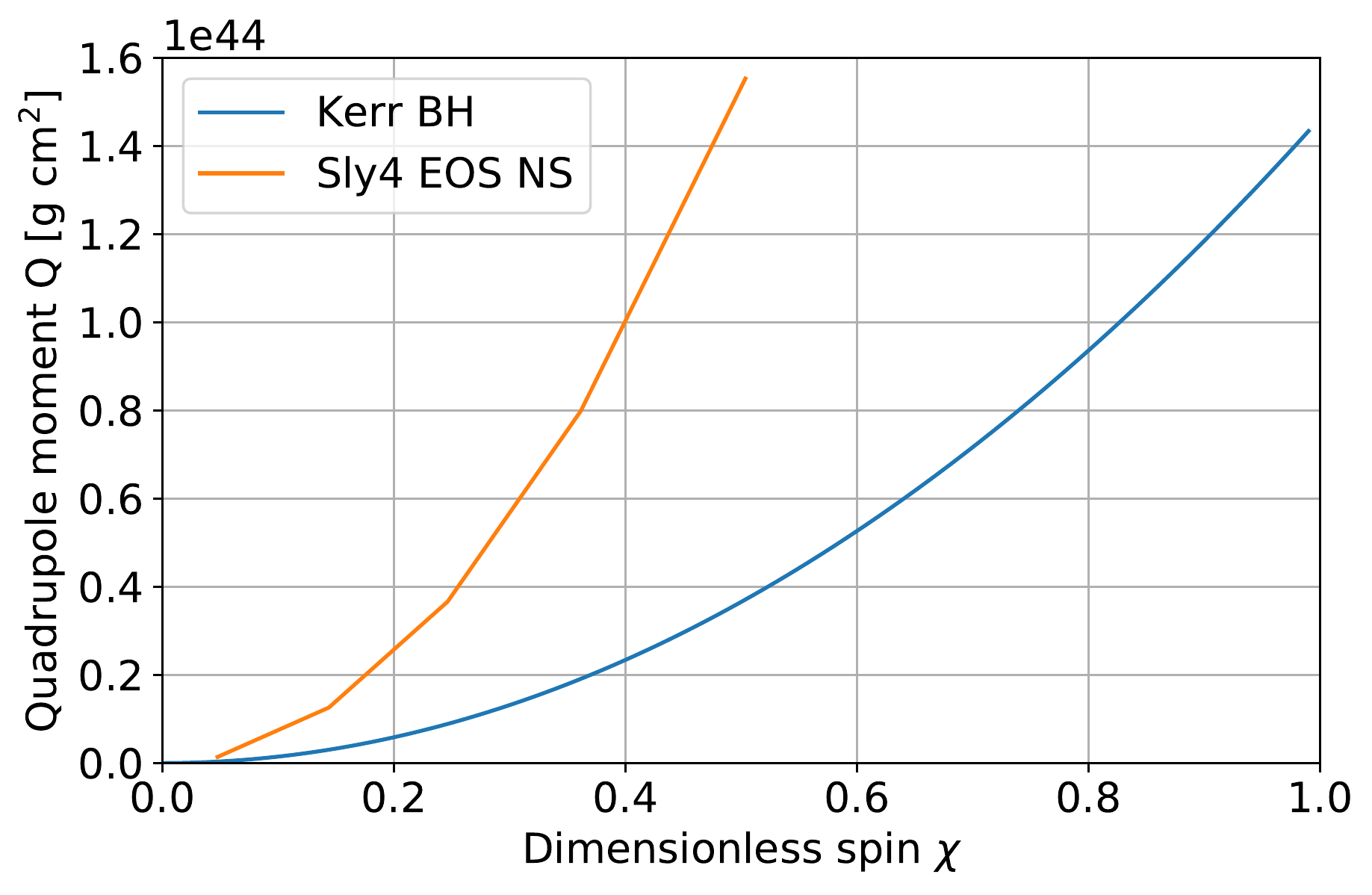} 
	\caption{Quadrupole moment $Q$ of a realistic EOS NS (Sly4 EOS of \citealt{Douchin2001}), 
and the Kerr BH with the same mass $M$ (equal to $M_N$ along the curves), 
as a function of the dimensionless spin parameter $\chi=ac^2/(GM)$.}
	\label{q-spin}
\end{figure}

The NS$\to$BH conversion and related NS spin-up alters the quadrupole moment 
of the object from a spin-dominated quadrupole moment of a NS,
%-----------------------------------------------------------------                            
\begin{equation}
  Q_{NS} \approx \mbox{$10^{43}$\ g cm$^2$}, 
\label{eq-quadrupole-NS}
\end{equation}
%-----------------------------------------------------------------   
(for a typical NS of mass $M=M_N$ rotating at $500$ Hz), 
to a well-defined quadrupole moment of a Kerr BH (see, e.g. \citealt{Misner1973}),
%-----------------------------------------------------------------                            
\begin{eqnarray}
  Q_{Kerr} &=& M a^2 =  \left({G}/{c^2}\right)^2 M^3 \chi^2 \nonumber \\  
  &\approx& 10^{44}\times \left({M}/{M_N}\right)^3\chi^2\ \mbox{g cm$^2$},  
\label{eq-quadrupole-Kerr}
\end{eqnarray}
%-----------------------------------------------------------------  
where $a=J/(Mc)$ denotes the BH spin, $\chi=ac^2/(GM)$ the dimensionless spin,  
and $J$ the total angular momentum. Quadrupole moments $Q_{NS}$ and $Q_{Kerr}$ 
for the same $M=M_N$ are plotted in Fig.~\ref{q-spin} as functions of $\chi$. 

In the following we will assume that the difference of $\Delta Q$ in the 
NS$\to$BH transition is of the order of $10^{43}$ g cm$^2$.  
%-----------------------------------------------------------------------
%-----------------------------------------------------------------------
Depending on the initial PBH mass and hence on the timescale of conversion, the
process  will be accompanied by the emission of gravitational waves with
frequencies in the planned spaceborne laser interferometer 
LISA\footnote{{\tt https://www.elisascience.org}} 
band, between $10^{-4}$ Hz and $10^{-1}$ Hz (see
\citealt{Cornish2018} for a sensitivity curve) for $M_P>10^{24}$ g. In general,
gravitational-wave energy emission due to the change of the quadrupole moment
$Q_{ij}$ in time $T$ can be approximated from the quadrupole formula
\citep{Einstein1918} as 
%-----------------------------------------------------------------                            
\begin{multline}
  \frac{dE}{dt} = \frac{G}{5c^5} \left<\dddot{Q}_{ij} \dddot{Q}^{ij} \right> 
  \approx \frac{G}{5c^5} \left(\frac{\Delta Q}{T^3}\right)^2\\    
  \approx 5\times 10^{25} \left(\frac{\Delta Q}{10^{43}\ \mbox{g cm}^2}\right)^2 
  \left(\frac{1\ \mbox{s}}{T}\right)^6 \mbox{erg/s}.    
\label{eq-dedt}
\end{multline}
%-----------------------------------------------------------------   
The instantaneous gravitational-wave amplitude $h_0$ may be approximated 
in the similar way as 
%-----------------------------------------------------------------                            
\begin{multline}
  h_0 = \frac{2G}{c^4 r} \ddot{Q}_{ij} 
  \approx \frac{2G}{c^4 r} \left(\frac{\Delta Q}{T^2}\right)\\
  \approx 5\times 10^{-34} 
  \left(\frac{\mbox{1 Gpc}}{r}\right) 
  \left(\frac{\Delta Q}{10^{43}\ \mbox{g cm}^2}\right)
  \left(\frac{1\ \mbox{s}}{T}\right)^2.  
\label{eq-h0}
\end{multline}
%-----------------------------------------------------------------
The above considerations suggest that even if the timescales of the 
conversion are of the order of the NS dynamical timescale, $10^{-3}$ s,  
the resulting strain amplitude is too small to be detectable at the  
cosmological distances. 

The Advanced LIGO and Advanced Virgo observations of the GW170817 event
\citep{Abbott2017} report a NS-mass binary system merger accompanied by a short
gamma-ray burst and a subsequent kilonova. The estimated rate or such events is
$1540^{+3200}_{-1220}$ Gpc$^{-3}$ y$^{-1}$ \citep{Abbott2017} which roughly
correspond to 1\% of the core-collapse supernova event rate $K^{N}_{tot}$. The observation 
does not provide a conclusive answer whether GW170817 merger consisted of two NS, or 
a NS and a light (NS mass) BH \citep{Abbott2017,Abbott2018,Yang2018}. 

The PBH-NS collisions provide a natural creation channel for light BH. 
To estimate a fraction of NS-mass binary system mergers containing a light BH we 
will use a simple rate estimate from Sect.~\ref{subsect:rate-analytic}. For 
the reference PBH mass $m_0$ we get from Eq.~\ref{frequency-single}, for a typical galaxy 
$\dot{n}\sim 3\times 10^{-6}\ \mathrm{y}^{-1}$. If the GW170817 merger rate 
corresponds to $\dot{n}_{GW}\sim 10^{-2}\times K^{N}_{tot}\sim 10^{-4}\ \mathrm{y}^{-1}$, 
then the fraction of merging binaries containing a light BH should be of the 
order of $\dot{n}/\dot{n}_{GW}\sim 3\%$.

%%%%%%%%%%%%%%%%%%%%%%%%%%%%%%%%%%%%%%%%%%%
\section{FRB duration} \label{sect:duration} 
%%%%%%%%%%%%%%%%%%%%%%%%%%%%%%%%%%%%%%%%%%%

The duration of an FRB event naturally stems from the 
characteristic NS mass and size timescale, which influences the reconnection 
processes in the NS magnetosphere. NS dynamical time is 
$\Delta t \approx r_G/c \approx 2GM_{\rm N}/c^3$. Thus, 
%-------------------------------------------------------------------------------------------------------
\begin{equation}
\delta t_{\rm model} \approx 10^{-3}\ \mbox{s}.
\label{duration-model}
\end{equation}
%-------------------------------------------------------------------------------------------------------   
This universal value does not depend on the PBH mass.

%%%%%%%%%%%%%%%%%%%%%%%%%%%%%%%%%%%%%%%%%%%%
\section{Discussion} \label{sect:discussion} 
%%%%%%%%%%%%%%%%%%%%%%%%%%%%%%%%%%%%%%%%%%%%

We present the analysis of a FRB creation scenario in which the dark-matter halo 
PBH from a mass range of $10^{25}\,\mbox{g}\gtrsim m \gtrsim
10^{17}\,\mbox{g}$ collide with galactic NS, nest at their centers and accrete
the dense matter converting the NS into a NS-mass BH. 

The FRB event itself is powered by the reconnecting NS magnetic field. 
Typical NS magnetic field (surface field of the order of $10^{12}$ G) stores
enough energy to provide for several FRB.

Depending on the timescale of Bondi-like accretion of dense matter onto the
PBH, which for initially small-mass PBH ($M_P < 10^{20}$ g) may take years,
gradual accretion of magnetic field confined in the superfluid dense matter
provides a source of repeating non-periodic FRB, explaining the FRB121102
observations. The ratio of the number of repeating FRB to non-repeating ones
may indicate that the mass function of PBH is dominated by heavier masses.  

Some FRB show a double peak structure: two bursts separated by a
millisecond-length interval. In the NS$\to$BH conversion model, the duration of
the interval is naturally associated with the rotation period of the NS being
spun-up in the process of BH formation in its center. 

Numerical simulations of the evolution of galactic PBH and NS populations were performed. For the PBH fraction of $0.01$, allowed by current observational constraints, we find that the mean PBH mass of order of $10^{21} $ g yields a rate of FRB throughout the Universe and the number of NS in the Galaxy consistent with observations. Simulations indicate qualitative differences between galactic disk and bulge populations of NS, depending on the mean PBH mass. Presented numerical model is rudimentary and can only be trusted to provide the order of magnitude results accuracy. Employing more sophisticated model for relevant parameters, such as NS creation rate, will allow for more specific predictions in the future. 

Presented model predicts a population of light, NS-mass BH. From the
micro-lensing observations point of view, this population is indistinguishable
from galactic NS. However, NS-mass BH may also be present in the relativistic
binary systems. Recent LIGO-Virgo observations of the binary NS-mass merger
GW170817 does not exclude the possibility of one component being a light BH.
Using the merger rate obtained by the LIGO and Virgo collaborations, based on
these recent observations, we estimate the fraction of events containing a
light BH to be of the order of a few per cent of the whole binary NS merger
population. 

We also estimate the change of the quadrupole moment of the object and related
gravitational-wave emission during the NS$\to$BH conversion. Our approximate
analysis suggests that the frequency of gravitational waves for a wide range of
PBH masses lies within the planned spaceborne LISA detector sensitivity range,
but the characteristic gravitational-wave amplitude is too small to be detected
from cosmological distances. 

Detailed analysis of the gravitational-wave background from this type of
sources and its detectability by the current and planned detectors is beyond
the scope of this paper, and will be addressed in a future work.    
  
\acknowledgments
This research was partially supported by the National Science Foundation under
grant NSF PHY-1125915 and by the Polish NCN grants 2014/14/M/ST9/00707,
2015/19/B/ST9/01099 and 2016/22/E/ST9/00037. MAA acknowledges the Czech Science
Foundation grant No. 17-16287S which supported one of his visits to the Paris
Observatory. We benefited from discussions with Ramesh Narayan, Pawe{\l}
Haensel, Avi Loeb, Axel Brandenburg and Krzysztof Belczy{\'n}ski. Some of this
work was conducted at the Black Hole Initiative at Harvard University, which is
funded through a grant from the John Templeton Foundation.



\end{document}